\documentclass[aps,twocolumn,showpacs,preprintnumbers,amsmath,amssymb,nofootinbib]{revtex4}
\usepackage{amssymb}
\usepackage{epsfig}
\usepackage{slashed}
\usepackage{graphicx}
\usepackage{bm}
\usepackage{color}

\begin{document}

\title{The NLO QCD corrections to $B_c$($B_c^*$) production around the $Z$ pole at an $e^+e^-$ collider}

\author{Xu-Chang Zheng$^{a,c}$\footnote{e-mail:zhengxc@itp.ac.cn}, Chao-Hsi Chang$^{a,b,c}$\footnote{e-mail:zhangzx@itp.ac.cn}, Tai-Fu
Feng$^{a,d}$\footnote{email:fengtf@hbu.edu.cn}, Zan Pan$^{a,c}$\footnote{e-mail:panzan@itp.ac.cn}}
\affiliation{$^a$ Key Laboratory of Theoretical Physics, Institute of Theoretical Physics, Chinese Academy of Sciences, Beijing 100190, China}
\affiliation{$^b$ CCAST (World Laboratory), P.O.Box 8730, Beijing 100190 China}
\affiliation{$^c$ School of Physical Sciences, University of Chinese Academy of Sciences, Beijing 100049, China}
\affiliation{$^d$ Department of Physics, Hebei University, Baoding, 071002, China}

\begin{abstract}
The production of $B_c$ and $B_c^*$ mesons at $Z$-factory (an $e^+e^-$ collider running at energies around the $Z$ pole) is calculated
up-to the next-to-leading order (NLO) QCD corrections. The results show that the dependence of the total cross sections on the renormalization scale $\mu$ is suppressed by the corrections, and the NLO corrections enhance the total cross sections for $B_c$ by $52\%$ and for $B_c^*$ by $33\%$, when the renormalization scale is taken at $\mu=2m_b$. To observe the various behaviors of the production of the mesons $B_c$ and $B_c^*$, such as the differential cross section vs. the out-going angle, the forward-backward asymmetry and the distribution vs. the energy fraction $z$ up-to QCD NLO accuracy as well as the relevant $K$-factor (NLO to LO) for the production are computed and it is pointed out that some of the observables obtained here may be used as specific precision test of the Standard Model.\\

\end{abstract}

\maketitle

\section{Introduction}
\label{intro}

The meson $B_c$ (and its anti-particle $\bar{B_c}$), being an explicitly heavy-flavored
quark-antiquark ground bound-state, is unique in the Standard Model (SM). The two components
inside it move non-relativistically due to heavy masses of its components,
so the potential model can describe the spectrum of the
binding system quite reliably\cite{pot}, and the nonrelativistic quantum chromodynamics effective theory (NRQCD)\cite{nrqcd} may
be adopted to compute its production, and with the effective theory for the weak interaction which is based on SM
its decays may be computed\cite{decays,ybook1,ybook2}, thus it specially interests us, particularly, since
it was observed by CDF collaboration firstly\cite{cdf}.

Since the observations on the meson $B_c$ (and its anti-particle $\bar{B_c}$) are available only at high energy hadronic colliders so far, so the theoretical and experimental studies of the $B_c$ production mostly focus on its hadronic production\cite{ybook1,ybook2}. According to QCD factorization theorem the hadronic production of a hadron, e.g. $B_c$ meson, always is through the collision of the partons inside the colliding hadrons stochastically, while the momentum fraction of the colliding hadron, carried by the colliding partons, is determined by the parton distribution function (PDF) of the colliding hadron, so the total colliding energy and the moving in longitudinal direction of the center-of-mass system (C.M.S.) of the colliding partons cannot be controlled, thus only the perpendicular components of the momenta of the products, which are measurable, have proper meaning in understanding the production. Namely to observe the production through hadron collision can acquire quite restrictive knowledge about the production.

In contrary, for the production of the meson $B_c$ (and its anti-particle $\bar{B_c}$) via $e^+e^-$ collisions, the C.M.S. of the 'subprocess' precisely is of $e^++e^-$ collisions, so the observables, such as all components of the momenta of the products, the angle distributions and the 'forward-backward asymmetry' of the concerned product to the direction of the colliding $e^+$ or $e^-$ etc, have proper meaning in studying the production, even may be used to test of the Standard Model, thus to study $B_c$ meson production at an $e^+e^-$ collider is very important and interesting. Especially the collisions happen to take place at a $Z$-factory ($e^+e^-$ colliders running at energies around the $Z$ pole), the production will be enhanced greatly by the resonance effect. Now several suggestions on $Z$-factories, $e^+e^-$ facilities run at energies around the $Z$ pole with much higher luminosity than that of LEP-I, e.g. ILC, CEPC and FCC-ee, are proposed, thus at a modernized $Z$-factory with very very high luminosity the production must achieve a lot of new knowledge, although the $B_c$ meson production at LEP-I (an old $Z$-factory) is too small to be observed\cite{changchen,lep-1}. Indeed concerning the possible $Z$-factory being under consideration, in Ref.\cite{doublyhadron} the production of doubly heavy flavored hadrons ($B_c$ meson and baryons $\Xi_{cc}, \Xi_{bc}, \Xi_{bb}$ etc and their excited states as well as their antiparticle) via $e^+e^-$ collision at the energy around the $Z$ pole is re-studied but only under the approach of complete QCD at the leading order (LO) and the fragmentation approach at the leading logarithm order (LL) thoroughly. In Ref.\cite{doublyhadron} it is found that the LO results have quite remarkable dependence on the renormalization scale, although certain interesting results, such as the precise asymmetries in forward-backward and lefthand-righthand in $B_c(B_c^*)$ production, are obtained. In order to have more precise theoretical prediction and to suppress the dependence on the renormalization scale, it certainly is requested to carry out the computations on the $B_c,(B_c^*)$ meson production at a $Z$-factory up-to the QCD NLO accuracy. Thus we devote ourselves to doing it here.

Since $B_c$ and its excited states such as $B_c^*,B_c^{**}\cdots$ carry two heavy-flavors explicitly, so the excited states $B_c^*,,B_c^{**}\cdots$ will decay (or cascade-decay) to the ground state, $B_c$, through strong or electromagnetic interaction with almost $100\%$ probability, thus as
Ref.\cite{doublyhadron}, here the production of the excited state $B^*_c$ ($^3S_1, J^P=1^-$), the lowest excited state of $B_c$, is
also computed up-to QCD NLO.

According to NRQCD\cite{nrqcd}, the production of $B_c(B_c^*)$ meson by electron-positron collision can be factorized into
two factors at a specific energy $\mu_F$ in QCD perturbative region: one is the electron-positron production of the
'free' $c\,,\bar{b}$ quark pair inclusively in short distance, which can be calculated by perturbative QCD (pQCD), and the other one is to depict how the produced $c$ and $\bar{b}$ quarks to form the meson $B_c(B_c^*)$, which is nonperturbative but can be achieved phenomenologically or via potential model (the wave function at origin) etc. Here setting the factorization energy scale $\mu_F$ is equal to the renormalization one $\mu_R$, for the former factor we compute the production up-to next leading order (NLO) QCD corrections, and for the later factor we consider the leading order in relative velocity $v$ between the two heavy quarks inside the meson $B_c(B_c^*)$ only. For convenience, later on we denote $\mu\equiv \mu_F=\mu_R$ throughout the paper.

The paper is organised as follows: Following the Introduction, in Section II, we briefly recall the useful formulas to the LO
accuracy. In Section III, we present the approaches to compute the NLO corrections of QCD for the $B_c$ (and $B_c^*$) meson production at
a $Z$-factory. In Section IV, with the necessary parameters being given, the numerical results
are presented. Section V is devoted to discussions and summary. In Appendix-A, it is shown how the relevant width of the production of $B_c$ meson and $B_c^*$ meson by $Z$ decay are derived from the total cross sections of the production at a $Z$-factory and in Appendix-B precise comparisons between the relevant widths derived from the total cross sections of the production at a $Z$-factory computed here and directly computed from the $Z$ decay, which appear in literature.

\section{The cross section up-to leading order (LO)}

\begin{figure}[htbp]
\includegraphics[width=0.45\textwidth]{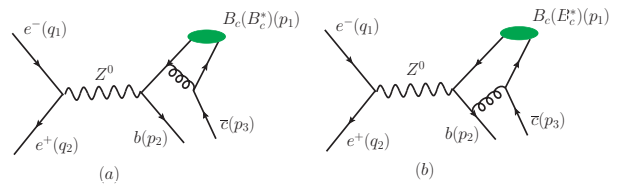}
\caption{Two of the four Feynman diagrams to LO accuracy for
the production $e^-(q_1)+e^+(q_2) \to B_c(B_c^*)(p_1) + b(p_2) +\bar{c}(p_3)$. } \label{feylo}
\end{figure}

There are four Feynman diagrams for the $B_c$ and $B_c^*$ production at LO accuracy, only two of them are presented in
Fig.\ref{feylo}, but the other two can be obtained by interchanging the $b-$quark and $c-$quark lines in Fig.\ref{feylo}.

Note that in the present paper the studies focus on the production from QCD LO accuracy up-to QCD NLO accuracy, but merely
when the $e^++e^-$ collider runs around the $Z$ pole. Thus the contributions corresponding to the Feynman diagrams with
$Z$-boson mediation are dominant and we will compute them carefully, but those corresponding to the Feynman diagrams with
$\gamma$ mediation are approximately ignored\footnote{Thus without special statement, in the Feynman diagrams Figs.\ref{feylo},\ref{feyct},\ref{feyself},\ref{feytri},\ref{feybox},\ref{feyreal} the $Z$ mediation diagrams
are involved only, but the ones with the $\gamma$ mediation are not.}. Under the approximation the computations for QCD LO and
QCD NLO are simplified quite a lot, and at the end of Section IV we also estimate how well the approximation is by taking into account the contributions from the Feynman diagrams with $\gamma$ mediation, i.e. those from the $\gamma$ mediation itself being squared and those
corresponding to the interference of the $\gamma$ mediation and $Z$-boson mediation.

The cross section to QCD LO can be formulated as:
\begin{equation}
d\sigma_{_{\rm{LO}}}= \frac{1}{4}\frac{1}{2s} \sum |M_{_{\rm{LO}}}|^{2} d\Phi_3 ,
\end{equation}
where $\frac{1}{2s}$ is the flux factor; $\sum$ means that the spins and the colors in the initial and final states are summed over; 1/4 comes from the spin average of the initial $e^+e^-$; $d\Phi_3$ denotes the three-body phase space for the final states and $M_{_{\rm{LO}}}$ is the LO Feynman amplitude, which is the sum of four terms for the LO Feynman diagrams. The details about $M_{_{\rm{LO}}}$ can be found in Ref.\cite{doublyhadron}.

\section{The NLO QCD corrections}
\label{NLOcalc}

The NLO QCD corrections for the process $e^++e^- \rightarrow B_c(B_c^*) + b +\bar{c}$ include virtual and real ones. The half of the Feynman diagrams for the virtual correction are those in Figs.\ref{feyct},\ref{feyself},\ref{feytri},\ref{feybox}, and the half of the Feynman diagrams for the real correction are those in Fig.\ref{feyreal}. The other half of the Feynman diagrams for the virtual corrections and for real corrections can be also obtained by interchanging the $b$-quark and $c$-quark lines in Figs.\ref{feyct}$\sim$\ref{feyreal}.

To the QCD NLO accuracy, the cross section is formulated as
\begin{equation}\label{NLOcalc0}
\sigma_{_{\rm{NLO}}}= \sigma_{_{\rm{LO}}}+\sigma_{\rm Virtual}+\sigma_{\rm Real}.
\end{equation}
Here $\sigma_{\rm Virtual}$ denotes the so-called virtual correction and $\sigma_{\rm Real}$ denotes the so-called real correction. Now let us calculate them respectively.

\begin{widetext}

\begin{figure}[htbp]
\includegraphics[width=0.70\textwidth]{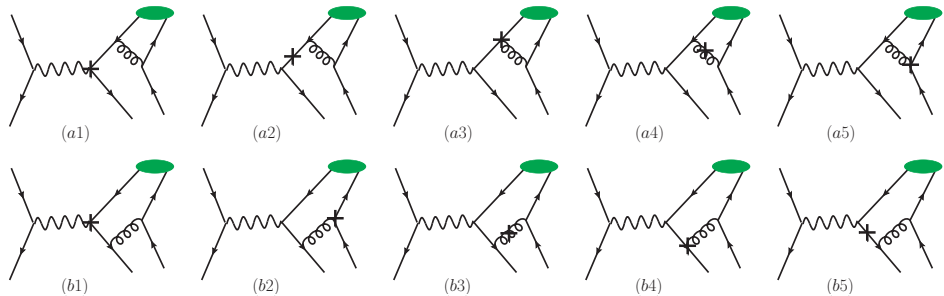}
\caption{The half of Feynman diagrams containing a `counterterm' (denoted by $\times$) in need of consideration for the NLO QCD
production $e^-+e^+ \rightarrow B_c(B_c^*) + b +\bar{c}$.} \label{feyct}
\end{figure}

\begin{figure}[htbp]
\includegraphics[width=0.60\textwidth]{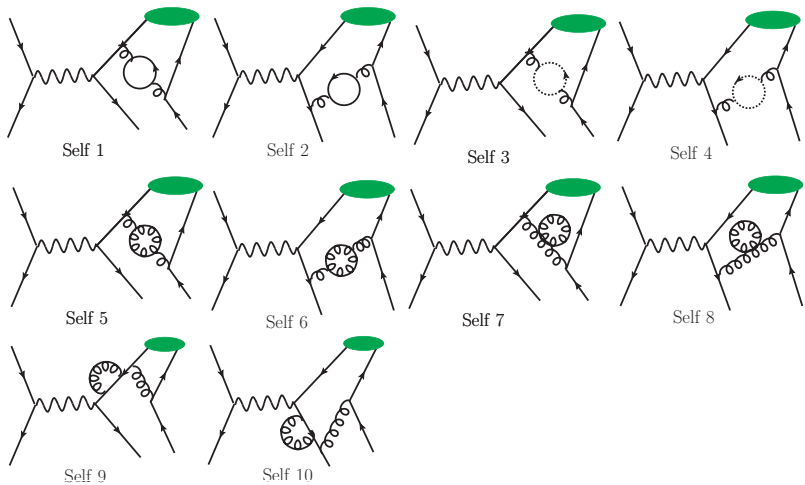}
\caption{The half of Feynman diagrams containing a `self-energy' part for the virtual correction in need of computation for
the NLO QCD production $e^-(q_1)+e^+(q_2) \to B_c(B_c^*)(p_1) + b(p_2) +\bar{c}(p_3)$.} \label{feyself}
\end{figure}

\begin{figure}[htbp]
\includegraphics[width=0.60\textwidth]{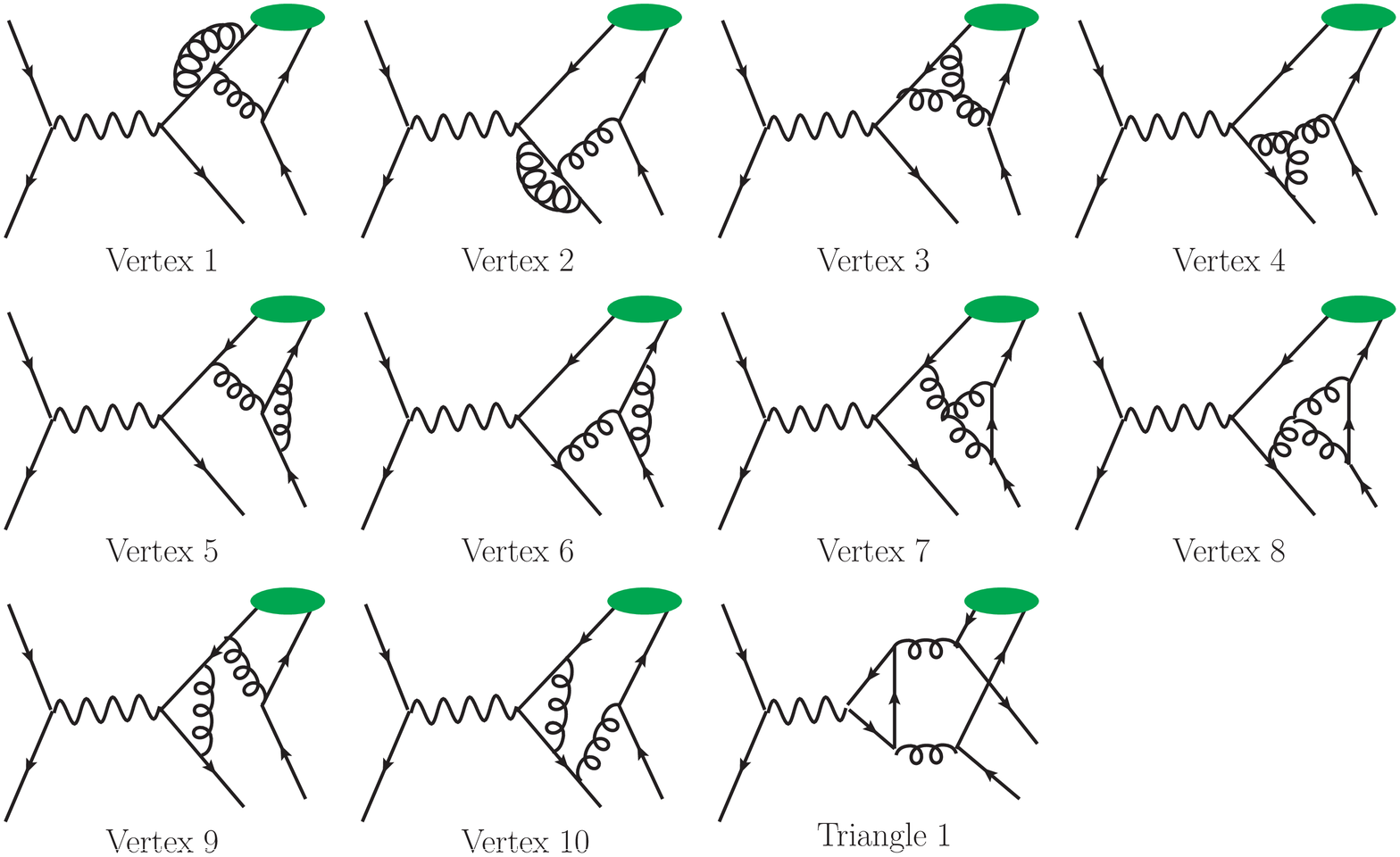}
\caption{The half of Feynman diagrams containing a vertex or a triangle part for the virtual correction in need of computation for
the NLO QCD production $e^-(q_1)+e^+(q_2) \to B_c(B_c^*)(p_1) + b(p_2) +\bar{c}(p_3)$. } \label{feytri}
\end{figure}

\begin{figure}[htbp]
\includegraphics[width=0.60\textwidth]{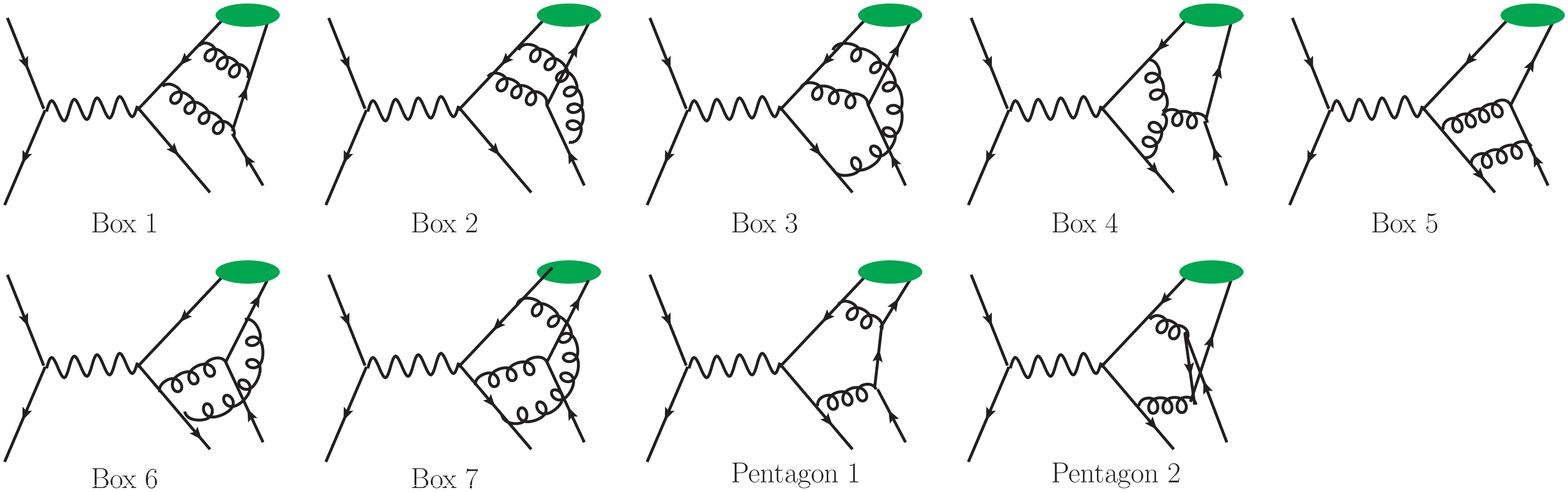}
\caption{The half of Feynman diagrams containing a `box' or a `pentagon' part for the virtual correction in need of computation for
the NLO QCD production $e^-(q_1)+e^+(q_2) \to B_c(B_c^*)(p_1) + b(p_2) +\bar{c}(p_3)$. } \label{feybox}
\end{figure}

\begin{figure}[htbp]
\includegraphics[width=0.60\textwidth]{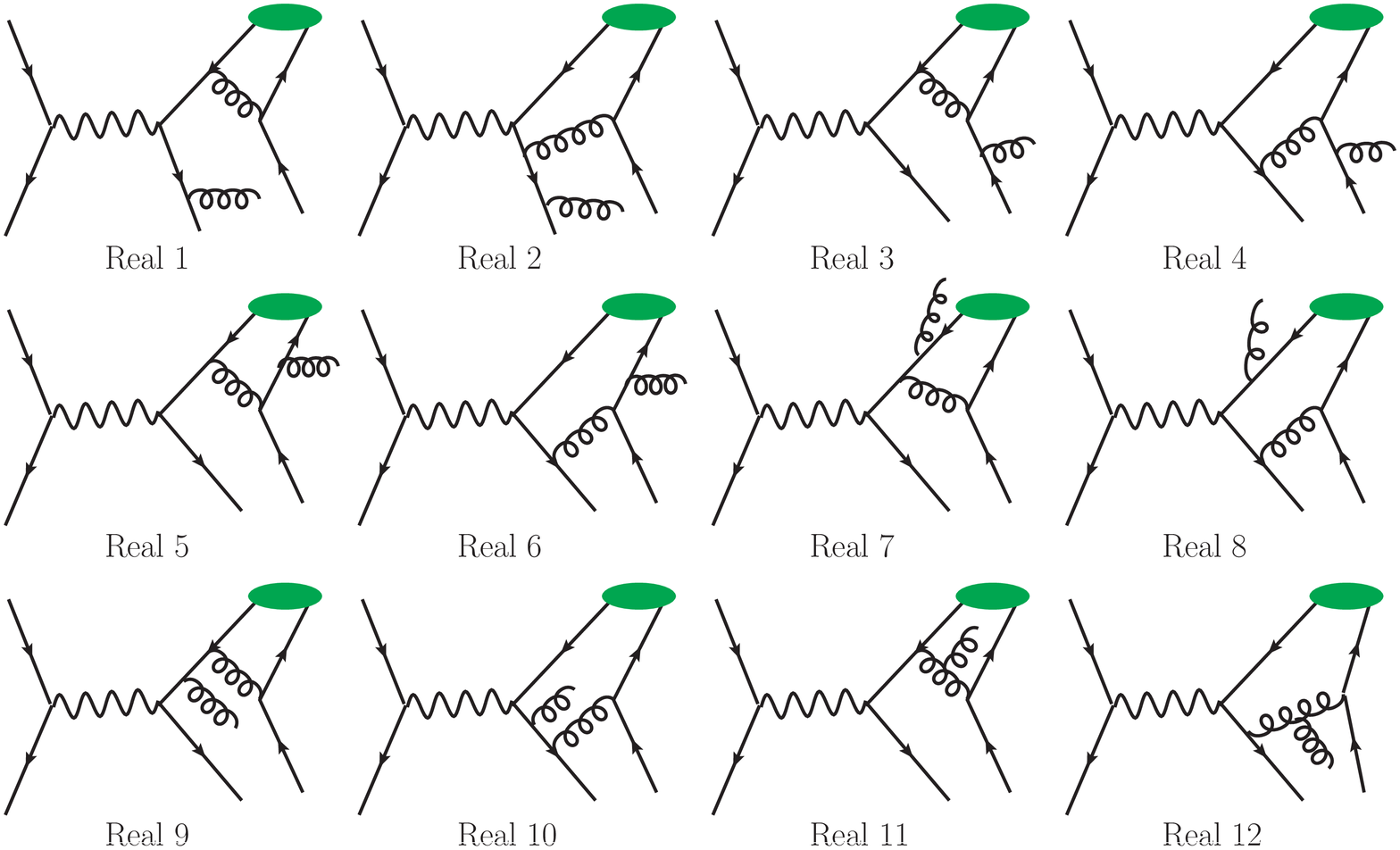}
\caption{The half of the Feynman diagrams for the real correction in need of computation for
the NLO QCD production $e^-(q_1)+e^+(q_2) \to B_c(B_c^*)(p_1) + b(p_2) +\bar{c}(p_3)$.} \label{feyreal}
\end{figure}

\end{widetext}

\subsection{The NLO QCD virtual correction}
\label{virtual}

The virtual correction up-to QCD NLO is to consider the interference of the LO ones and those corresponding to the correction
Feynman diagrams, i.e. Figs.\ref{feyct},\ref{feyself},\ref{feytri},\ref{feybox} and those with
the $c$-quark and $b$-quark lines being interchanged. Thus the virtual correction to the cross section can be formulated as
\begin{equation}\label{virtual01}
d\sigma_{\rm Virtual}= \frac{1}{4}\frac{1}{2s} \sum 2 {\rm Re} (M^*_{_{\rm{LO}}} M_{\rm Virtual}) d\Phi_3.
\end{equation}


There are ultraviolet (UV) and infrared (IR) divergences in the amplitudes corresponding to the correction Feynman diagrams.
We adopt dimensional regularization with $D=4-2\epsilon$ to isolate the UV and IR divergences. There are the Coulomb divergences in the conventional matching procedure, which should be absorbed into the binding potential for the two heavy quarks inside the $B_c$ and $B_c^*$. In the dimensional regularization, there is a simpler way to extract the NRQCD short-distance coefficients directly using the method of regions\cite{region}, i.e., expanding the amplitudes with the relative momentum ($q$) of the constituent quarks before performing loop integration, and in the lowest non-relativistic approximation for the $S$-wave states of the binding system $c\bar{b}$, only the terms with $q=0$ are taken. Thus we don't confront the contributions from the low energy regions such as those from the potential region.

In dimensional regularization, $\gamma_5$ should be treated carefully. We adopt the reading point
prescription\cite{gamma5}. It has the following rules,
\begin{itemize}
\item The anticommutation relation $\lbrace \gamma_5,\gamma^{\mu} \rbrace=0$ is valid. Thus after applying
the anticommutation relation and $\gamma^2_5=1$, there is one or no $\gamma_5$ in each Dirac trace.
\item Cyclic manipulation in the Dirac traces is prevented. When considering the contributions from several diagrams,
for all of them the traces in the amplitude (or resulting from squared fermionic amplitudes) must be read with starting
from the same vertex respectively.
\item The relevant axial current anomalies would be obtained and the conservation for vector currents is guaranteed
by starting all the traces with the axial vector vertex.
\end{itemize}

Here the UV divergences come from self-energy, vertex and triangle diagrams only \footnote{The UV divergence from the amplitude of the anomalous diagram Triangle-1 in Fig.\ref{feytri} is canceled by the UV divergence from the other anomalous diagram.}, which are canceled by the counterterms through renormalization, and here the renormalization scheme is that the renormalization constants $Z_2$, $Z_m$, and $Z_3$, which correspond to the renormalization of quark field, quark mass and gluon field, are determined by the renormalization of the on-mass-shell scheme (OS), whereas $Z_g$ relating to the strong coupling constant $\alpha_s$ is determined by the renormalization of the modified-minimal-subtraction scheme ($\overline{MS}$). Then with the renormalization, we have:
\begin{eqnarray}
\label{rencont}
\delta Z^{OS}_2&=&-C_F \frac{\alpha_s}{4\pi}\left[\frac{1}{\epsilon_{UV}}+ \frac{2}{\epsilon_{IR}}-3~\gamma_E+3~ {\rm ln}\frac{4\pi \mu^2}{m^2}+4\right], \nonumber\\
\delta Z^{OS}_m&=&-3~C_F \frac{\alpha_s}{4\pi}\left[\frac{1}{\epsilon_{UV}}- \gamma_E+
 {\rm ln}\frac{4\pi \mu^2}{m^2}+\frac{4}{3}\right],\nonumber\\
 \delta Z^{OS}_3&=&\frac{\alpha_s}{4\pi}\left[(\beta'_0-2C_A)\left(\frac{1}{\epsilon_{UV}}-\frac{1}{\epsilon_{IR}}\right) \right. \nonumber\\
 &&\left.-\frac{4}{3}T_F \left(\frac{1}{\epsilon_{UV}}-\gamma_E + {\rm ln}\frac{4\pi \mu^2}{m_c^2}\right)\right. \nonumber\\
 &&\left.-\frac{4}{3}T_F \left(\frac{1}{\epsilon_{UV}}-\gamma_E + {\rm ln}\frac{4\pi \mu^2}{m_b^2}\right)\right], \nonumber\\
 \delta Z^{\overline{MS}}_g&=&- \frac{\beta_0}{2}\frac{\alpha_s}{4\pi}\left[\frac{1}{\epsilon_{UV}}- \gamma_E+ {\rm ln}~(4\pi) \right],
\end{eqnarray}
where $m$ appearing in $\delta Z^{OS}_2$ and $\delta Z^{OS}_m$ represents the mass $m_b$ or $m_c$ accordingly, $\mu$ is the energy where the renormalization is carried out, and $\gamma_E$ is Euler's constant. $\beta_0=\frac{11}{3}C_A-\frac{4}{3}T_F n_f$ is the one-loop coefficient of the QCD $\beta$-function, and $n_f$ is the number of active quark flavors. Here for the concerned process, there are three light quarks $u,d,s$ and two heavy quarks $c,b$, so $n_f=5$. But in Eq.(\ref{rencont}) precisely $\beta'_0=\frac{11}{3}C_A-\frac{4}{3}T_F n_{lf}$, and $n_{lf}=3$ for the light quark flavors.
For $SU(3)_c$ group, $C_A=3$, $T_F=\frac{1}{2}$ and $C_F=\frac{4}{3}$. Because there is no external gluon line at LO level, $\delta Z_3$ is canceled at NLO total amplitude level, so the final results are independent of the renormalization scheme of the gluon field.


The IR divergences in the Feynman diagrams for the virtual correction can be well analyzed\cite{irdiv1,irdiv2}. For the concerned process, the IR divergences come from the vertex , box and pentagon diagrams. Of Fig.\ref{feytri}, only the amplitudes corresponding to diagrams Vertex-5 and Vertex-6 have IR divergences. Of Fig.\ref{feybox}, except Box-4, the rests have IR divergences. The other half of the Feynman diagrams, which are obtained from Figs.\ref{feytri}$\sim$\ref{feybox} by interchanging the $c$-quark and $b$-quark lines, have similar IR divergences. The IR divergences in the virtual correction will be canceled by the IR divergences from the counterterms and the real correction.

\subsection{The real corrections to NLO}
\label{real}

Note that here `the NLO real correction'\footnote{In literature, sometimes `the NLO real correction' contains only the contributions from the process $e^+e^-\rightarrow B_c(B_c^*)+b+\bar{c}+g$ with the gluon $g$ so soft or collinear to merge into $b$ or $\bar{c}$ jet.} for the concerned process $e^-+e^+\to B_c(B_c^*)+b+\bar{c}$
means to take into account the full contributions from the process $e^-(q_1)+e^+(q_2)\to B_c(B_c^*)(p_1)+b(p_2)+
\bar{c}(p_3)+g(p_4)$ with an additional gluon in final
state but covering whole possible phase space.

Half of the Feynman diagrams for the real correction are shown in Fig.\ref{feyreal} and the other half can be obtained from Fig.\ref{feyreal} through interchanging the $c$-quark and $b$-quark lines. The correction to the relevant cross section can be written as:
\begin{equation}
d\sigma_{\rm Real}= \frac{1}{4}\frac{1}{2s} \sum |M_{\rm Real}|^{2} d\Phi_4,
\end{equation}
where $M_{\rm Real}$ is the sum of 24 terms relating to the 24 Feynman diagrams for the real correction. $|M_{\rm Real}|^{2}$ can be formulated as
\begin{equation}
|M_{\rm Real}|^{2}= \sum_{i,j} M^*_{\rm Real,i}M_{\rm Real,j},
\end{equation}
where $i,j$ vary from 1 to 24 corresponding to the 24 real correction Feynman diagrams.

There are IR divergences in the real correction, which are generated by the phase space integration, and they should be finally canceled by the IR divergences appearing in the virtual correction. It is easy to realize\cite{irreal} that, the terms $M^*_{\rm Real,i}M_{\rm Real,j}$ are IR finite for the phase space integration unless $M_{\rm Real,i}$ and $M_{\rm Real,j}$ are the amplitudes corresponding to Feynman diagrams in which a real gluon is emitted from an external on-shell line, e.g., here the first 8 diagrams in Fig.\ref{feyreal}. At this step, let us divide the cross section of the real correction into two parts as
\begin{equation}\label{NLOcalc01}
d\sigma_{\rm Real}=d\sigma^{\rm IR}_{\rm Real}+d\sigma^{\rm IR-finite}_{\rm Real},
\end{equation}
where $d\sigma^{\rm IR}_{\rm Real}$ contains the terms in $\sum_{i,j} M^*_{\rm Real,i}M_{\rm Real,j}$ only when $M_{\rm Real,i}$ and $M_{\rm Real,j}$ are the amplitudes corresponding to Feynman diagrams in which a real gluon is emitted from an external on-shell line. Namely we can formulate $d\sigma^{\rm IR}_{\rm Real}$ as
\begin{eqnarray}
d\sigma^{\rm IR}_{\rm Real}= \frac{1}{4}\frac{1}{2s} \sum |M^{\rm IR}_{\rm Real}|^{2} ~d\Phi_4
\end{eqnarray}
where $M^{\rm IR}_{\rm Real}$ is the sum of the amplitudes corresponding to Feynman diagrams in which a real gluon is emitted from an external
on-shell line. $d\sigma^{\rm IR-finite}_{\rm Real} $ contains the remaining terms in $\sum_{i,j} M^*_{\rm Real,i}M_{\rm Real,j}$, i.e., $(|M_{\rm Real}|^{2}- |M^{\rm IR}_{\rm Real}|^{2})$. Due to the fact that there is no divergence in $\sigma^{\rm IR-finite}_{\rm Real} $, we can calculated it in 4-dimensional space-time directly.

In order to extract the IR divergences in $\sigma^{\rm IR}_{\rm Real}$ precisely so as
to cancel the IR divergences in virtual correction precisely, we adopt the two-cutoff phase space slicing method\cite{twocut}.
By this method, the integration on the phase space is divided into two sectors through introducing a very soft cut $\delta_s(\ll1)$
on the energy of the emitting gluon ($p_4^0$). Then,
\begin{equation}
\label{1o1}
d\sigma^{\rm IR}_{\rm Real}=d\sigma^{\rm IR,soft}_{\rm Real}+d\sigma^{\rm IR,hard}_{\rm Real} ,
\end{equation}
where
\begin{eqnarray}
\label{1o2}
d\sigma^{\rm IR,hard}_{\rm Real}= \frac{1}{4}\frac{1}{2s} \sum |M^{\rm IR}_{\rm Real}|^{2} ~d\Phi_4 \vert _{p_4^0>\delta_s\sqrt{s}/2},
\end{eqnarray}
and
\begin{eqnarray}
\label{1o3}
d\sigma^{\rm IR,soft}_{\rm Real}= \frac{1}{4}\frac{1}{2s} \sum |M^{\rm IR}_{\rm Real}|^{2} ~d\Phi_4 \vert _{p_4^0<\delta_s\sqrt{s}/2}.
\end{eqnarray}

To calculate $\sigma^{\rm IR,soft}_{\rm Real}$, the eikonal approximation is adopted to deal with the amplitudes involved in $M^{\rm IR}_{\rm Real}$, where the terms of ${\cal O}(\delta_s)$ in $\sigma^{\rm IR,soft}_{\rm Real}$ have been neglected \cite{twocut,eikapp}. Under this approximation, the amplitude corresponding to the Feynman diagram where a real gluon emitted from an external line can be factorized as a Born factor multiplying an eikonal factor, and it is easy to check that the eikonal factors relating to the diagrams Real-5 and Real-6 of Fig.\ref{feyreal} are canceled by the eikonal factors relating to the diagrams Real-7 and Real-8 of Fig.\ref{feyreal} respectively at the leading order approximation in relative velocity $\mathcal{O}(v^0)$. Thus at last under the eikonal approximation we obtain
\begin{widetext}
\begin{equation}
\sum |M^{\rm IR}_{\rm Real}|^{2}=4 \pi \alpha_s C_F \mu^{2\epsilon}\left[\frac{-(p_2)^2}{(p_2.p_4)^2}+\frac{2p_2.p_3}{(p_2.p_4)(p_3.p_4)}-\frac{(p_3)^2}{(p_3.p_4)^2}\right]\sum |M_{\rm Born}|^{2}\,.
\label{MIRsoft}
\end{equation}
Up-to corrections of ${\cal O}(\delta_s)$, the phase space for the soft sector can be factorized as \cite{twocut}
\begin{equation}
d\Phi_4 \vert _{p_4^0<\delta_s\sqrt{s}/2}=d\Phi_3 \frac{d^{d-1}p_4}{2p_4^0 (2\pi)^{d-1}} \vert _{p_4^0<\delta_s\sqrt{s}/2}
\label{SPsoft}
\end{equation}
where $d\Phi_3$ denotes the element of the three-body phase space without emitting a gluon. Performing the integration over the momentum of the emitting gluon ($p_4$) in the soft sector, the differential cross section
\begin{equation}
d\sigma^{\rm IR,soft}_{\rm Real}= d\sigma_{\rm Born}\frac{\alpha_s}{\pi}\Gamma(1+\epsilon)\left( \frac{4\pi\mu^2}{s}\right)^{\epsilon}\left( \frac{A}{\epsilon}+B\right)
\label{IRsoft}
\end{equation}
is obtain\cite{softint}, where
\begin{eqnarray}
A&=& C_F \left[1-\frac{\kappa~ (p_2\cdot p_3)}{\kappa^2~ m_b^2-m_c^2} {\rm ln}\left( \frac{\kappa^2~ m_b^2}{m_c^2}\right) \right], \nonumber \\
B&=&C_F\left\lbrace -\left[1-\frac{\kappa~ (p_2\cdot p_3)}{\kappa^2~ m_b^2-m_c^2} {\rm ln}\left( \frac{\kappa^2~ m_b^2}{m_c^2}\right) \right]{\rm ln}(\delta_s^2)+\frac{1}{2\beta_b}{\rm ln}\left( \frac{1+\beta_b}{1-\beta_b}\right)+\frac{1}{2\beta_{\bar{c}}}{\rm ln}\left( \frac{1+\beta_{\bar{c}}}{1-\beta_{\bar{c}}}\right)\right.\nonumber\\
&&\left. + \frac{2\kappa~ (p_2\cdot p_3)}{\kappa^2~ m_b^2-m_c^2}\left[ \frac{1}{4}{\rm ln}^2\left( \frac{u^0-\vert \textbf{u} \vert}{u^0+\vert \textbf{u} \vert}\right)+{\rm Li}_2\left( 1-\frac{u^0+\vert \textbf{u} \vert}{v}\right)  +{\rm Li}_2\left( 1-\frac{u^0-\vert \textbf{u} \vert}{v}\right)   \right]^{u=\kappa p_2}_{u=p_3}  \right\rbrace ,
\end{eqnarray}
\end{widetext}
here
\begin{eqnarray}
\beta_b&=&\sqrt{1-m_b^2/(p_2^0)^2},\nonumber\\
\beta_{\bar{c}}&=&\sqrt{1-m_c^2/(p_3^0)^2},
\end{eqnarray}
\begin{equation}
v=\frac{\kappa^2~m_b^2-m_c^2}{2(\kappa~p_2^0-p_3^0)},
\end{equation}
and
\begin{equation}
\kappa=\frac{p_2\cdot p_3+\sqrt{(p_2\cdot p_3)^2-m_b^2~m_c^2}}{m_b^2}\,.
\end{equation}
It can be checked precisely that the $1/\epsilon$-terms in $d\sigma^{\rm IR,soft}_{\rm Real}$ defined by Eq.(\ref{IRsoft}) are
just canceled by those infrared $1/\epsilon$-terms remained by the virtual correction Eq.(\ref{virtual01}).

Since there is no IR divergence in $d\sigma^{\rm IR,hard}_{\rm Real}$ due to the constraint $p_4^0>\delta_s\sqrt{s}/2$, so we can calculate it in 4-dimensional space-time safely. Summing up $\sigma^{\rm IR-finite}_{\rm Real}$, $\sigma^{\rm IR,soft}_{\rm Real}$ and $\sigma^{\rm IR,hard}_{\rm Real}$,  the requested $\sigma_{\rm Real}$ is obtained. Then with the Eqs.(\ref{NLOcalc0},\ref{virtual01},\ref{NLOcalc01}), the cross section $\sigma_{_{NLO}}$ of the process $e^++e^-\to B_c(B_c^*)+ b +\bar{c}+X$, i.e. the production to QCD NLO accuracy,
is achieved.

\section{Numerical Results}
\label{numer}

For numerical calculations, the necessary input parameters are taken as follows:
\begin{eqnarray}
& &m_b=4.9 ~{\rm GeV}\,,\; m_c=1.5 ~{\rm GeV}\,,\; m_{_Z}=91.1876~ {\rm GeV}\,,\nonumber \\
& & {\rm sin^2\theta_w}=0.231\,,\; \alpha=1/128\,,\; \Gamma_{_Z}=2.4952~{\rm GeV}\,, \nonumber \\
& &\vert R_S(0) \vert^2=1.642 ~{\rm GeV^3}\,,
\end{eqnarray}
$\alpha=\alpha(m_{_Z})$ is the electromagnetic coupling constant at $\mu=m_{_Z}$; $R_S(0)$ is the radial wave function at the origin for $B_c$($B_c^*$), which can be taken from the potential model\cite{pot}. We apply the two-loop formula for the strong coupling constant $\alpha_s(\mu)$:
\begin{equation}
\alpha_s(\mu)=\frac{4\pi}{\beta_0~{\rm ln}(\mu^2/\Lambda^2_{QCD})}\left[ 1-\frac{\beta_1~{\rm ln}~{\rm ln}(\mu^2/\Lambda^2_{QCD})}{\beta_0^2~{\rm ln}(\mu^2/\Lambda^2_{QCD})}\right],
\end{equation}
where $\beta_1=\frac{34}{3}C_A^2-4C_F T_F n_f-\frac{20}{3}C_A T_F n_f$ is the two-loop coefficient of the QCD $\beta$-function. According to $\alpha_s(m_{_Z})=0.1185$\cite{PDG}, we obtain $\Lambda^{nf=5}_{QCD}=0.233$ GeV.

\begin{table}[h]
\begin{tabular}{c c c c c}
\hline\hline
~$\mu$~  & ~~$\alpha_s(\mu)$~~ & ~~$\sigma_{\rm LO}({\rm pb})$~ & ~$\sigma_{\rm NLO}({\rm pb})$~ & ~$K\equiv \sigma_{\rm NLO}/\sigma_{\rm LO}$~  \\
\hline
$2m_b$ & ~0.180~ & ~1.576~ & 2.387 & 1.515 \\
$m_{_Z}/2$ &~0.132~ & ~0.847~ & 1.587 & 1.874 \\
\hline\hline
\end{tabular}
\caption{The total cross section of $e^+e^- \to B_c+b+\bar{c}+X$ at the $Z$ pole with two typical renormalization scales.}
\label{bcsection}
\end{table}

\begin{widetext}

\begin{table}
\begin{tabular}{c c c c c}
\hline\hline
~$\mu$~  & ~~$\alpha_s(\mu)$~~ & ~~$\sigma_{\rm LO}({\rm pb})$~ & ~$\sigma_{\rm NLO}({\rm pb})$~ & ~$K\equiv \sigma_{\rm NLO}/\sigma_{\rm LO}$~  \\
\hline
$2m_b$ & ~0.180~ &  ~2.204~ & 2.930 & 1.329 \\
$m_{_Z}/2$ &~0.132~ & ~1.185~ & 2.059 & 1.738 \\
\hline\hline
\end{tabular}
\caption{The total cross section of $e^+e^- \to B_c^*+b+\bar{c}+X$ at the $Z$ pole with two typical renormalization scales.}
\label{bcsection2}
\end{table}

\begin{table}
\begin{tabular}{ c c c c c c c c c c }
\hline\hline
~~$cos\theta$~~  & ~~-0.8~~ & ~~-0.6~~ & ~~-0.4~~   & ~~-0.2~~   & ~~0~~  & ~~0.2~~    & ~~0.4~~   & ~~0.6~~   & ~~0.8~~  \\
\hline
$d\sigma/dcos\theta (B_c,LO)$     & 1.066   & 0.892 & 0.759 &0.667  &0.617 &0.608 & 0.639 & 0.711 & 0.825     \\
$d\sigma/dcos\theta (B_c,NLO)$    & 1.606   & 1.346 & 1.150 &1.014  &0.939 &0.924 & 0.969 & 1.075 & 1.242    \\
$K (B_c) $    & 1.506   & 1.509 & 1.515 &1.520  &1.522 &1.520 & 1.516 & 1.512 & 1.505    \\
$d\sigma/dcos\theta (B_c^*,LO)$   & 1.507   & 1.254 & 1.060 &0.926  &0.853 &0.839 & 0.884 &0.990 & 1.156  \\
$d\sigma/dcos\theta (B_c^*,NLO)$   & 1.990   & 1.662 & 1.414 &1.240  &1.144 &1.125 & 1.183 & 1.317 & 1.529    \\
$K (B_c^*)$   & 1.320   & 1.325 & 1.334 &1.339  &1.341 &1.341 & 1.338 & 1.330 & 1.323   \\
\hline\hline
\end{tabular}
\caption{The LO and NLO differential cross sections $\frac{d\sigma}{d~cos\theta}$ (in $pb$) of $e^-e^+ \to B_c(B_c^*)+b+\bar{c}+X$ and their ratio at various scattering angles ($\cos\theta$) at the $Z$ pole ($\mu=2m_b$).}
\label{tcos}
\end{table}

\begin{table}
\begin{tabular}{c c c c c c c c c c c }
\hline\hline
~~$z$~~  & ~~0.183~~ & ~~0.269~~ & ~~0.355~~ & ~~0.441~~   & ~~0.527~~   & ~~0.613~~  & ~~0.699~~    & ~~0.785~~   & ~~0.871~~   & ~~0.957~~   \\
\hline
$d\sigma/dz (B_c,LO)$  & 0.276  & 0.543   & 0.833 & 1.195 &1.655  &2.237 &2.932 & 3.603 & 3.664 & 1.534     \\
$d\sigma/dz (B_c,NLO)$  & 0.650  & 1.173   & 1.682 & 2.274 &2.964  &3.732 &4.508 & 4.970 & 4.360 & 1.578   \\
$K (B_c) $   & 2.355  & 2.160   & 2.019 & 1.903 &1.791  &1.668 &1.538 & 1.379 & 1.190 & 1.029    \\
$d\sigma/dz (B_c^*,LO)$  & 0.167  & 0.417   & 0.699 & 1.091 &1.681  &2.582 &3.905 & 5.584 &6.617 & 3.187   \\
$d\sigma/dz (B_c^*,NLO)$ & 0.446  & 0.920   & 1.418 & 2.029 &2.884  &4.098 &5.586 & 7.056 & 7.058 & 2.852   \\
$K (B_c^*) $ & 2.671  &2.206   & 2.029 & 1.860 &1.716  &1.587 &1.430 & 1.264 & 1.067 & 0.895    \\
\hline\hline
\end{tabular}
\caption{The LO and NLO differential cross sections $\frac{d\sigma}{dz}$ (in $pb$) of $e^-e^+ \to B_c(B_c^*)+b+\bar{c}+X$ and their ratios vs. various values of $z$ (the energy fraction carried by $B_c(B_c^*)$) at the $Z$ pole peak ($\mu=2m_b$).}
\label{tz}
\end{table}

\begin{table}[h]
\begin{tabular}{c c c c c c c c c c c c c c}
\hline\hline
~~$(\sqrt{s}-m_{_Z})$({\rm GeV})~~  & ~~-5~~ & ~~-2.5~~ & ~~-1.5~~ & ~~-0.8~~   & ~~-0.4~~   & ~~-0.2~~  & ~~0~~    & ~~0.2~~   & ~~0.4~~   & ~~0.8~~ & ~~1.5~~ & ~~2.5~~ & ~~5~~ \\
\hline
$B_c$ (LO)  & 0.09  & 0.30   & 0.63 & 1.10 &1.42  &1.53 &1.58 & 1.54 & 1.44 & 1.13 & 0.66 & 0.32 & 0.10   \\
$B_c$ (NLO)  & 0.13  & 0.46   & 0.95 & 1.66 &2.13  &2.32 &2.39 & 2.33 & 2.18 & 1.71 & 1.00 & 0.49 & 0.15   \\
$B_c^*$ (LO)  & 0.12  & 0.42   & 0.88 & 1.54 &1.98  &2.14 &2.20 & 2.16 & 2.02 & 1.59 & 0.92 & 0.46 & 0.14   \\
$B_c^*$ (NLO) & 0.16  & 0.56   & 1.17 & 2.04 &2.64  &2.84 &2.93 & 2.87 & 2.67 & 2.11 & 1.23 & 0.60 & 0.18   \\
\hline\hline
\end{tabular}
\caption{The total cross sections (in $pb$ and with $\mu=2m_b$) of $e^-e^+ \to B_c(B_c^*)+b+\bar{c}+X$ at the collision energies around $m_{_Z}$ ($Z$-boson peak).}
\label{bcEcm}
\end{table}



\begin{table}
\begin{tabular}{c c c c c c c c c c c c c c}
\hline\hline
~~$(\sqrt{s}-m_{_Z})$({\rm GeV})~~  & ~~-5~~ & ~~-2.5~~ & ~~-1.5~~ & ~~-0.8~~   & ~~-0.4~~   & ~~-0.2~~  & ~~0~~    & ~~0.2~~   & ~~0.4~~   & ~~0.8~~ & ~~1.5~~ & ~~2.5~~ & ~~5~~ \\
\hline
$B_c$  & 0.18  & -0.35   & -0.63 & -0.53 &0.06  &0.33 &0.79 & 1.24 & 1.64 & 2.11 & 2.21 & 1.91 & 1.39   \\
$B_c^*$  & 0.19  & -0.53   & -0.92 & -0.78 &-0.12  &0.41 &1.04 & 1.67 & 2.22 & 2.88 & 3.01 & 2.60 & 1.88   \\
\hline\hline
\end{tabular}
\caption{The contributions (in $fb$ and with $\mu=2m_b$) to the production due to the Feynman diagrams with a photon mediation to replace the $Z$-boson mediation at the collision energies around $m_{_Z}$.}
\label{gamma}
\end{table}

\end{widetext}

\begin{figure}[htbp]
\includegraphics[width=0.3\textwidth]{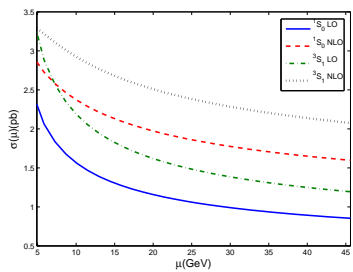}
\caption{The dependence of the total cross sections for $B_c(B_c^*)$ production on the renormalization scale $\mu$ at the $Z$ pole to LO and NLO.} \label{mu}
\end{figure}

Note that in the calculations here, we use FeynArts\cite{feynarts} for generating Feynman diagrams and amplitudes, FeynCalc\cite{feyncalc} and FeynCalcFormlink\cite{formlink} for carrying out the trace of color and Dirac matrices; while Apart\cite{apart} and FIRE\cite{fire} for conducting partial fraction and integration-by-parts (IBP) reduction. All the one-loop integrals are reduced into master integrals and the master integrals are computed in terms of LoopTools\cite{looptools} numerically. The final phase-space integrations are computed with the help of
the soft-ware Vegas\cite{vegas}.

The numerical results of the total cross sections at the colliding center-mass energy $m_Z$ as well as the so-called $K$-factor (QCD) for the productions， $e^+e^- \to B_c+b+\bar{c}+X$ and $e^+e^- \to B_c^*+b+\bar{c}+X$，with two different renormalization scales, $\mu=2m_b$ and $\mu=m_{_Z}/2$, are put into the tables: TABLE \ref{bcsection} and TABLE \ref{bcsection2} respectively, and the precise dependence of the cross sections on the renormalization scale $\mu$ for LO and NLO QCD is presented in Fig.\ref{mu}. It is shown in Fig.\ref{mu} that the cross section of the production $e^+e^- \to B_c (B_c^*)+b+\bar{c}+X$ at the $Z$ pole decreases by $46\%$($46\%$) at LO but by $34\%$ ($30\%$) at NLO when the renormalization scale $\mu$ changes from $2m_b$ to $m_{_Z}/2$. Namely the dependence on renormalization scale $\mu$ is weaken a lot due to NLO correction. Whereas the dependence on the renormalization scale $\mu$ is still quite great for NLO, thus it seems that, to suppress the dependence on $\mu$ further, higher order corrections in QCD for the concerned production are requested.

Moreover, we also with $\mu=2m_b$ calculate the differential cross sections $d\sigma/d\cos\theta$, $d\sigma/dz$ and the relevant $K$-factor as well. Here $\theta$ is the angle between the momenta of the electron in initial state and the meson $B_c$($B_c^*$) in final state at center-of-mass system of the $e^+,e^-$ collision and $z$ is the `energy-fraction' defined as $2k \cdot p_1/s$ ($k$ is the momentum carried by $Z$ boson).

\begin{figure}[htbp]
\includegraphics[width=0.3\textwidth]{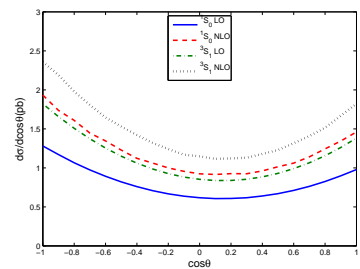}
\caption{The differential cross sections $\frac{d\sigma}{d~cos\theta}$ to LO and NLO 
for $e^-e^+ \to B_c(B_c^*)+b+\bar{c}+X$ at the $Z$ pole ($\mu=2m_b$).} \label{cos}
\end{figure}

The differential cross sections $d\sigma/d\cos\theta$ for the production of $B_c$($B_c^*$) meson with $\mu=2m_b$ to LO and NLO, and the $K$ factor as well are put in TABLE \ref{tcos}. The differential cross section $d\sigma/d\cos\theta$ with renormalization scale $\mu=2m_b$ is presented in Fig.\ref{cos}. It is shown by Fig.\ref{cos} that due to the NLO QCD corrections the differential cross section $d\sigma/d\cos\theta$ changes only within a common factor $K$ presented in TABLE \ref{tcos}.

From Fig.\ref{cos} the asymmetry in $d\sigma/d\cos\theta$ due to $Z$-boson mediation at the levels of LO and NLO can be seen very clear, which
varies with the values of the electroweak mixing angle ${\rm sin^2\theta_w}$ for $b$ and $c$ quarks. When measuring the asymmetry and to suppress the experimental systematic errors, similar to what done by LEP-I and SLC\cite{pdg}, one may introduce the forward-backward asymmetry $A_{FB}$ for the $B_c$ or $B_c^*$ production as follows, although LEP-I and SLC is for measuring the forward-backward asymmetry for heavy quarks and leptons and here is for measuring the forward-backward asymmetry for the $B_c(B_c^*)$ production which relates to the electroweak mixing angle $\sin\theta_W$ for $b$ and $c$ quarks
only:
\begin{equation}
A_{FB}=\frac{\sigma_F-\sigma_B} {\sigma_F+\sigma_B},
\end{equation}
where $\sigma_F$ denotes the cross section for $\theta \in (0,\pi/2)$ and $\sigma_B$ denotes the cross section for $\theta \in (\pi/2,\pi)$ thus
we compute the forward-backward asymmetry $A_{FB}$ for the production of $B_c$ and $B_c^*$ meson from LO to NLO:
\begin{eqnarray}
&A^{LO}_{FB}(B_c)=-9.58\times 10^{-2}\,,\\ \nonumber
&A^{NLO}_{FB}(B_c)=-9.50\times 10^{-2}\,, \\ \nonumber
&A^{LO}_{FB}(B_c^*)=-9.97\times 10^{-2}\,, \\ \nonumber
&A^{NLO}_{FB}(B_c^*)=-9.83\times 10^{-2}\,.
\end{eqnarray}
The forward-backward asymmetry $A_{FB}$ for the production of $B_c$ and $B_c^*$ meson is about ten percent, that is easy to be seen experimentally.

\begin{figure}[htbp]
\includegraphics[width=0.3\textwidth]{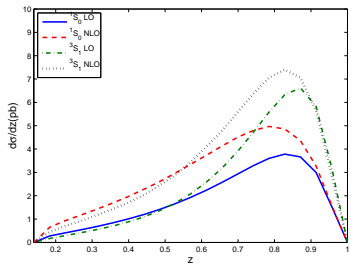}
\caption{The differential cross sections $\frac{d\sigma}{dz}$ to LO and NLO 
for $e^-e^+ \to B_c(B_c^*)+b+\bar{c}+X$ at the $Z$ pole peak ($\mu=2m_b$).} \label{z}
\end{figure}

The energy-fraction distributions of $B_c$ and $B_c^*$ production, $\frac{d\sigma}{dz}$, are also computed with $\mu=2m_b$, and the precise values obtained for the distributions are put into TABLE \ref{tz} and in Fig.\ref{z} for the relevant curves. One may see from TABLE \ref{tz} that the $K$ factors vary with the energy-fraction $z$ quite a lot, and from Fig.\ref{z} that the maximum point of the $z$ distributions is shifted to smaller $z$ due to QCD NLO corrections.

To be a reference and to see the variation of the total cross section at the collision energies around $m_{_Z}$ peak (within 5 GeV region), we
also compute the cross section to LO and NLO with $\mu=2m_b$, and put the result in TABLE.\ref{bcEcm}.

By the way, we should note here that for the production of $B_c$ and $B_c^*$ by $Z$ decay the relevant decay widths for $\Gamma_{Z \to B_c(B_c^*)+b+\bar{c}+X}$ can be easily `read out' from the total cross sections of the production, $\sigma(e^+e^- \to B_c(B_c^*)+b+\bar{c}+X)$, at the $Z$ pole peak with the contributions for photon mediation being ignored. In the literature there are the relevant decay widths, $\Gamma_{Z \to B_c(B_c^*)+b+\bar{c}+X}$, at LO and NLO\cite{qiao1,qiao2}, thus we have made precise comparisons of the widths read out from the total cross sections with those computed directly from the $Z$ decay in literature. In Appendix-A, the way how to 'read out' the decay widths $\Gamma_{Z \to B_c(B_c^*)+b+\bar{c}+X}$ respectively from the total cross sections $\sigma(e^++e^-\to B_c(B_c^*)+b+\bar{c}+X)$ at the $Z$ pole is presented, and in Appendix-B careful comparisons are made. The situation is that the induced NLO decay width $\Gamma_{Z \to B_c+b+\bar{c}+X}$ is bigger than that in Ref.\cite{qiao1}, but the induced NLO decay width $\Gamma_{Z \to B_c^*+b+\bar{c}+X}$ is consistent with that in Ref.\cite{qiao2}.



In the above calculations the contributions from the photon mediation are ignored. In order to see the ignored contributions the squared Feynman diagrams which have a photon mediation instead of a $Z$-boson mediation and the interference of the Feynman diagrams with a $Z$-boson mediation and those with a photon mediation should be computed. The two components: the amplitude with a photon mediation squared and the interference of those with a $Z$-boson mediation and with a photon mediation, and put the results in TABLE \ref{gamma}. From the results TABLE \ref{gamma} one may see that the contributions from the ones with a photon propagator (those squared and the interference) around the $Z$ pole resonance are very small in comparison with the contributions from those with a $Z$-boson propagator, thus it may be conclude that the approximation ignoring the contributions from the Feynman diagrams with a virtual $\gamma$ is quite good, furthermore since the precise values on the contributions estimated to QCD LO accuracy
in TABLE \ref{gamma} are so small, so it is reasonable to believe that the conclusion will be still valid even the estimate on the contributions up-to QCD NLO accuracy.

\section{Discussions and conclusion}
\label{conclusion}

We have calculated the NLO QCD corrections to the production of $B_c$ or $B_c^*$ meson at $e^+e^-$ colliders
running near the $Z$ pole. The results show that the NLO corrections are significant. The dependence on the
renormalization scale $\mu$ for the cross sections at NLO level is suppressed in comparison with LO results.
Precisely, the total cross section for the $B_c$ ($B_c^*$) production at the $Z$-pole peak decreases about $46\%$ ($46\%$)
for LO, and about $34\%$ ($30\%$) for NLO when $\mu$ changes from $2m_b$ to $m_{_Z}/2$ accordingly. Namely, the dependence
of the cross sections on the renormalization scale $\mu$ up-to NLO correction is still not small, so it
means that to suppress the $\mu$ dependence further higher order QCD corrections are requested.

According to the present NLO QCD calculations, the conclusion obtained by LO calculations keeps valid.
Namely to study the production $e^+e^- \to B_c (B_c^*)+b+\bar{c}+X$ experimentally in order to enhance the
statistics of the relevant events the best energy region is around the $Z$ pole peak for resonance enhancement
and the collider still is requested to have so high luminosity i.e. higher than ${\rm 10^{35}~cm^{-2}s^{-1}}$,
because up-to QCD NLO accuracy the cross sections for the production do not change much, i.e. still
are of the order $\mathcal{O}(pb)$.

The computations and analyses here on the total cross sections, the differential cross sections vs. angle,
the forward-backward asymmetry and energy fraction distributions of the produced $B_c$ and $B^*_c$ mesons show that the shape
of the angle distributions, the forward-backward asymmetry, the $K$-factor from the NLO QCD accuracy to the LO QCD accuracy change
slightly, but the distribution on the energy fraction $z$ changes sizable i.e. the maximum is shifted to smaller energy fraction (see TABLE. \ref{tcos}, Figs. \ref{cos} and \ref{z}). Therefore, when experimentally the events have been collected numerously enough
(it is accessible for a collider with so high luminosity as mentioned above), the characters in angle distributions and
the forward-backward asymmetry etc for the production up-to NLO accuracy may be used not only to test of the theoretical
predictions for the production but also as done by LEP-I and SLC to do the precision test of SM, e.g. to test the electro-weak
mixing angle ${\rm sin^2\theta_w}$ etc.

\vspace{4mm}

\noindent {\bf\Large Acknowledgments:} We thank Jian-Xiong Wang and Rong Li for helpful discussions. This work was supported
in part by Nature Science Foundation of China (NSFC) under Grant No. 11275243, No. 11275036, No. 11447601, No. 11535002 and
No. 11675239

\vspace{4mm}
\centerline{\bf\large Appendix}
\vspace{-6mm}
\subsection{The decay width $\Gamma(Z\to B_c(B_c^*)+b+\bar{c}+X)$ reduced from the total cross section $\sigma(e^+e^-\rightarrow B_c(B_c^*)+b+\bar{c}+X)$ at the energy of the $Z$-boson pole}

The total cross section of the process $e^+e^-\rightarrow `Z' \rightarrow B_c(B_c^*)+b+\bar{c}+X$ can be represented as
\begin{equation}
\sigma=\frac{1}{4}\frac{1}{2s}\frac{L^{\mu\nu}H_{\mu\nu}}{(s-m_{_Z}^2)^2+m_{_Z}^2 \Gamma_{_Z}^2}\,,
\label{a1}
\end{equation}
where $\Gamma_Z$ is the total width of the boson $Z$, $L^{\mu\nu}$ is the leptonic tensor and
\begin{equation}
L^{\mu\nu}=a\left[ q_1^{\mu}q_2^{\nu}+q_2^{\mu}q_1^{\nu}-(s/2)g^{\mu\nu}\right]+i b \epsilon^{\mu\nu\lambda\tau}q_{1\lambda}q_{2\tau}\,,
\label{a2}
\end{equation}
here
\begin{eqnarray}
a&=&\frac{e^2(1-4{\rm sin}^2\theta_w+8{\rm sin}^4\theta_w)}{2{\rm sin}^2\theta_w{\rm cos}^2\theta_w}\,,\nonumber \\
b&=&\frac{e^2(1-4{\rm sin}^2\theta_w)}{2{\rm sin}^2\theta_w{\rm cos}^2\theta_w}\,.
\end{eqnarray}
$H_{\mu\nu}$ is the hadronic tensor, which has been performed the phase space integration. $H_{\mu\nu}$ can only depend on $k$, and has following form
\begin{equation}
H_{\mu\nu}=H_1(s)g_{\mu\nu}+H_2(s)k_{\mu}k_{\nu}/s\,,
\label{a3}
\end{equation}
where $H_1(s)$ and $H_2(s)$ are scalar functions. Because there's no UV or IR divergence after considering the renormalization and the real correction, $H_1(s)$ and $H_2(s)$ are free of UV and IR divergences. Thus, all the derivations in the appendix are performed in 4-dimension. According to Eqs.(\ref{a1}), (\ref{a2}) and (\ref{a3}), we can obtain the cross section at the $Z$ pole
\begin{equation}
\sigma=-\frac{a H_1(m_{_Z}^2)}{8m_{_Z}^2\Gamma_{_Z}^2}\,,
\label{a4}
\end{equation}
here $\sqrt s=m_{_Z}$.

The decay width of the process $Z\rightarrow B_c(B_c^*)+b+\bar{c}+X$ is
\begin{equation}
\Gamma=\frac{1}{3}\frac{1}{2m_{_Z}}\Pi^{\mu\nu}H_{\mu\nu}\,,
\label{a5}
\end{equation}
where
\begin{equation}
\Pi^{\mu\nu}=-g^{\mu\nu}+k^{\mu}k^{\nu}/m_{_Z}^2\,,
\label{a5}
\end{equation}
then we obtain
\begin{equation}
\Gamma=-\frac{H_1(m_{_Z}^2)}{2m_{_Z}}\,.
\label{a6}
\end{equation}

According to Eqs.(\ref{a4}) and (\ref{a6}), we obtain the relation
\begin{equation}
\Gamma=\frac{4m_{_Z}\Gamma_{_Z}^2}{a}\sigma\,.
\label{a7}
\end{equation}
Obviously from Eq.(\ref{a7}) one can obtain the decay width for $Z \to B_c(B_c^*)+b+\bar{c}+X$ from the total cross section of the relevant process $e^+e^-\to B_c(B_c^*)+b+\bar{c}+X$.


\subsection{To compare the results for the decay $Z\to B_c(B_c^*)+b+\bar{c}+X$ up-to NLO}

\begin{table}[h]
\begin{tabular}{c c c}
\hline\hline
~~~~$\mu$~~~~~& ~~$~~~~\Gamma_{\rm NLO}({\rm Ours})$~~~~ &~~~~$\Gamma_{\rm NLO}$(Ref.\cite{qiao1})~ \\
\hline
$2m_b$  & $111.05\pm0.09$ &~78.45 \\
$m_{_Z}/2$ & $76.22 \pm0.04$& ~62.53 \\
\hline\hline
\end{tabular}
\caption{The derived width (in keV) for the decay $Z \to B_c+b+\bar{c}+X$ from the total cross section for the relevant production $e^++e^-\to B_c+b+\bar{c}+X$ with the same input parameters as those in Ref.\cite{qiao1}. The values in the last column of the table are copied from Ref.\cite{qiao1}. Here for the derived NLO width, the statistical errors from the numerical integration on the phase space are also presented.}
\label{bcwidth}
\end{table}



\begin{table}[h]
\begin{tabular}{c c c}
\hline\hline
~~~~~$\mu$~~~~  & ~~~~$~~~\Gamma_{\rm NLO}(\rm Ours)$~~~~~ &~~~~$\Gamma_{\rm NLO}$(Ref.\cite{qiao2})~ \\
\hline
$2m_b$ & $118.48\pm0.09$& 118.77 \\
$m_{_Z}/2$  & $84.47\pm0.05$ & 84.60 \\
\hline\hline
\end{tabular}
\caption{The derived width (in keV) for the decay $Z \to B_c^*+b+\bar{c}+X$ from the total cross section for the relevant production $e^++e^-\to B_c^*+b+\bar{c}+X$ with the same input parameters as those in Ref.\cite{qiao2}, and the values in the last column of the table are copied from Ref.\cite{qiao2}. Here for the derived NLO width, the statistical errors from numerical phase space integration are also presented.}
\label{bcwidth2}
\end{table}


There are calculations on the decay width for $Z \to B_c(B_c^*)+b+\bar{c}+X$ up-to NLO QCD in Refs.\cite{qiao1,qiao2}, so here we do the comparisons on the decay width of theirs and those derived from our calculations for the total cross section of the production
$e^++e^-\to B_c(B_c^*)+b+\bar{c}+X$ in terms of the way in Appendix-A.

The two-cutoff phase space slicing method for the phase space integration by introducing $\delta_s$\cite{irdiv1} is used
as indicated by Eqs.(\ref{1o1},\ref{1o2},\ref{1o3}) and $\delta_s$ is fixed as $10^{-6}$ finally according to the requirement for the
method. So it would be better that the errors generated by the calculation are presented precisely for the comparisons.
Moreover, in order to compare the results in Refs.\cite{qiao1,qiao2} with ours, we take the same parameters as those taken
in Refs.\cite{qiao1,qiao2} and put the comparison results in TABLE \ref{bcwidth} and TABLE \ref{bcwidth2}.

From the tables, one may see that the results for NLO corrections of $B_c$ production in \cite{qiao1} are different from ours, but those for NLO corrections of $B^*_c$ production in \cite{qiao2} are consistent with ours.

\end{document}